\documentclass[aps,reprint,twocolumn,pre,showpacs,superscriptaddress,floatfix,nofootinbib]{revtex4-1}
\usepackage{amssymb,amsmath,amsfonts} \usepackage{epsfig,graphicx}
\begin{document}
\title{Irreversible processes without energy dissipation in an
  isolated Lipkin-Meshkov-Glick model}

  \author{Ricardo Puebla} \affiliation{Institut f\"{u}r Theoretische
    Physik, Albert-Einstein Allee 11, Universit\"{a}t Ulm, 89069 Ulm,
    Germany}\email{ricardo.puebla@uni-ulm.de} 
  \affiliation{Grupo de F\'isica Nuclear, Departamento de
    F\'isica At\'omica, Molecular y Nuclear, Universidad Complutense
    de Madrid, Av. Complutense s/n, 28040 Madrid, Spain}
   \author{Armando Rela\~{n}o}
  \affiliation{Departamento de F\'{\i}sica Aplicada I and GISC,
    Universidad Complutense de Madrid, Av. Complutense s/n, 28040
    Madrid, Spain}
\begin{abstract}
  For a certain class of isolated quantum systems, we report the
  existence of irreversible processes in which the energy is {\em not}
  dissipated. After a closed cycle in which the initial energy
  distribution is fully recovered, the expectation value of a
  symmetry-breaking observable changes from a value different from
  zero in the initial state, to zero in the final state. This entails
  the unavoidable loss of a certain amount of information, and
  constitutes a source of irreversibility. We show that the von
  Neumann entropy of time-averaged equilibrium states increases in the
  same magnitude as a consequence of the process. We support this
  result by means of numerical calculations in an experimentally
  feasible system, the Lipkin-Meshkov-Glick model.
\end{abstract}


\maketitle

\section{Introduction} 
A fundamental explanation of entropy production, irreversibility and
dissipation is the cornerstone of non-equilibrium statistical
mechanics. It is now well known that microscopic time reversibility
entails a number of relations between dissipation in any
time-dependent process and the thermodynamic properties of the
equilibrium initial and final states~\cite{Jarzynski:11}. For isolated
systems, the work dissipated in a cyclic process is linked to the
increase of entropy after the forward part, if the system equilibrates
to a microcanonical state after each time-dependent
process~\cite{Talkner:08}. But it is also stated that there is a
strong connection between irreversibility and information gained or
lost by the system, manifested in the physical consequences of the
Szilard engine~\cite{Szilard:29}, and expressed in the Landauer
principle of information erasure \cite{Landauer:61}. So, a complete
description of the entropy production following a thermodynamic
transformation requires the account of the information
flow~\cite{Sagawa:10}, and the addition of information reservoirs,
besides the traditional heat and chemical baths~\cite{Jordan:14}.

This is specially relevant in isolated quantum systems that
equilibrate to complex equilibrium states~\cite{Polkovnikov:11} which
store relevant information in sets of commuting constants of
motion~\cite{Rigol:07}, or in the coherences between different
subspaces of the system in the case of degenerate
spectra~\cite{Puebla:13,epl}. This means that the Hamiltonian $H$ and
the number of particles $N$ are not enough to characterize the
equilibrium states and the thermodynamic processes in this kind of
systems. Hence, a number of fundamental questions naturally arise. How
the extra information stored in the equilibrium state of the system is
eroded by an irreversible time-dependent protocol? Must it be
explicitly included in the formulation of the Second Law?

In the present article we deal with these questions in a certain class
of isolated quantum systems. We report the existence of processes for
which the energy is {\em not} dissipated, but they are irreversible
due to an unavoidable loss of information about the initial
symmetry-breaking. Hence, we conclude that this information has to be
accounted for a precise definition of irreversibility and for a
consistent formulation of the Second Law.

This phenomenon can be observed in systems that exhibit a transition
from a normal or non-degenerate, to a double-degenerate phase in the
energy spectrum. These systems have an extra discrete symmetry $S$
which labels all the eigenstates, although symmetry-breaking
eigenstates can also exist in the double-degenerate phase. After a
thermodynamic transformation leading the system from the
double-degenerate to the normal region, phase mixing between different
symmetry sectors~\footnote{The existence of a discrete symmetry $S$
  provides that the Hilbert space can be split in a direct sum of as
  many subspaces as different eigenvalues has the corresponding
  symmetry operator. Each of these subspaces is called symmetry
  sector.} forces the evolved state to be in a superposition of the
different branches of symmetry-breaking. As a consequence, the
expectation values of symmetry-breaking observables become zero at the
end of the protocol, independently on the initial condition; we can
thus conclude that the information about the initial symmetry-breaking
is lost at the end of the protocol. On the contrary, observables that
are not linked to symmetry-breaking, such as the energy, are not
affected; if the protocol is slow enough, the energy distribution is
exactly recovered at the end. So, an unexpected kind of
irreversibility arises, not related to energy dissipation, but to the
loss of significant information about the initial state.  We support
our conclusions with numerical calculations in an experimentally
feasible system~\cite{Zibold:10,Gross:10}.

The paper is organized as follows. In Sec. II we describe the model
and the protocol used to implement the irreversible processes. In
Sec. III we present the main numerical results. In Sec. IV we provide
an interpretation in terms of the von Neumann entropy. In Sec. V we
propose a mechanism which accounts for this unexpected irreversible
behavior. Finally, in Sec. VI we summarize the main conclusions.

\section{Model and protocol}

\subsection{Many-body quantum system} 

We illustrate this phenomenon in the Lipkin-Meshkov-Glick
model~\cite{Lipkin:65, Vidal:04} that describes $N$ interacting
particles of $\frac{1}{2}$-spin coupled to an external field, which
has been experimentally explored with a great level of control,
accuracy and isolation~\cite{Zibold:10}. In its most general version,
it includes both $J_x^2$ and $J_y^2$ interaction terms, and two free
parameters: one for the coupling strength and another for the
deformation~\cite{Vidal:04}. However, its most recent experimental
realization~\cite{Zibold:10} consists of a condensate of $N$ atoms
distributed between two different modes, whose interaction is governed
by the $J_x^2$ term. In this way, the model reduces to the following
Hamiltonian
\begin{equation} 
H = \chi J_x ^2 - \Omega J_z ,
\label{eq:ham}
\end{equation} 
where $\vec{J}$ is the Schwinger pseudospin representation of the $N$
two-level atom system; the parameters $\chi$ and $\Omega$ describe the
nonlinearity of the atom-atom interaction and the linear coupling
strength, respectively~\cite{Zibold:10}; $J_x$ gives the difference of
population between the two modes, and $J_y$ and $J_z$ the
corresponding coherences. Since $J^2$ is a conserved quantity, we
consider only the sector of maximum angular momentum,
i.e. $J=N/2$. This is precisely what has been measured in recent
experiments~\cite{Zibold:10,Gross:10}. The rest of sectors behave in
the same qualitative way; all are equivalent to the one with $J=N/2$,
but with different effective number of particles and very large
degeneracies, due to the many different possibilities of obtaining a
certain value of the angular momentum $J<N/2$ by coupling $N$
particles of $\frac{1}{2}$-spin.

The Hamiltonian~(\ref{eq:ham}) holds a discrete and global symmetry
$S$, which leads to another conserved quantity, the parity $\Pi=e^{i
  \pi \left(J_z+J\right)}$. Defining a new parameter, $\Lambda=
\frac{\chi N}{\Omega}$, and rescaling Eq. (\ref{eq:ham}) one can work
in terms of $\Lambda$, which is the external parameter that controls
the dynamics of the system. We consider $\hbar=1$ throughout this
article, thus $\chi$ and $1/\chi$ are the units of energy and time
respectively.

\begin{figure}
  \includegraphics[width=0.8\linewidth,angle=-0]{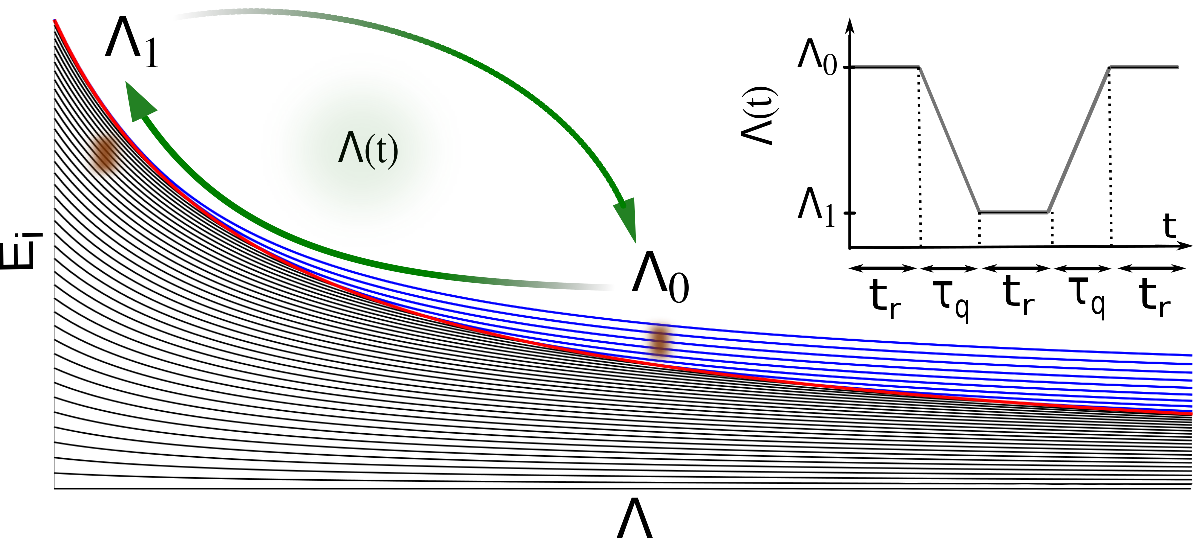}
  \caption{Scheme of the energy spectrum as a function of
    $\Lambda$. The critical line corresponding to $E_c$ is represented
    by a thick line (red online). Above, there is double-degenerate
    region (blue online); below, the normal region (black
    online). Arrows (green online) sketch the forward and the backward
    processes.  In the inset, the function $\Lambda(t)$ is shown.}
\label{fig:energy}
\end{figure}

The energy spectrum is divided in two regions (see
Fig. \ref{fig:energy}), one where eigenstates with opposite parity are
degenerate, i.e $H\left|E_i,\pi=\pm
\right\rangle=E_{i,\pm}\left|E_i,\pi=\pm\right\rangle$ with
$E_{i,+}=E_{i,-}$ being
$\Pi\left|E_i,\pi_n\right\rangle=\pi_n\left|E_i,\pi_n\right\rangle$,
and another without degeneracies. Hence, the Complete Set of Commuting
Observables (CSCO) of the Hamiltonian (\ref{eq:ham}) is $\{H, \Pi \}$,
though the Hamiltonian itself is enough to label all the eigenstates
in the non-degenerate region. The border between these two regions has
been identified as an excited-state quantum phase transition
(ESQPT)~\cite{Perez:08} and takes place at the critical energy
$E_c=2J^2/\Lambda$. If $E<E_c$ there are no degeneracies and every
eigenstate has a well-defined parity. As $\left \langle E_i,\pi_n
\right| J_x\left| E_i,\pi_m \right\rangle \propto
\delta_{\pi_n,-\pi_m}$, the expectation value of this observable in
any eigenstate of this part of the spectrum is zero. On the contrary,
if $E>E_c$ any combination of $\left|E_i,+ \right \rangle$ and $\left
|E_i,- \right \rangle$ is also an eigenstate of $H$. As a consequence,
this region is characterized by two symmetry-breaking branches, one
with $\left\langle J_x \right\rangle > 0$, and another with
$\left\langle J_x \right\rangle < 0$~\cite{Puebla:13,epl}. This is
very similar to what happens in many second-order phase transitions,
like in the Ising model ---below the critical temperature, two
branches of magnetization appear, and the spin-flip symmetry is
spontaneously broken.  So $\left<J_x \right>$ emerges as a good order
parameter for the ESQPT.

\subsection{Protocol} 

As it is sketched in Fig.~\ref{fig:energy}, we complete a cycle by
linearly changing $\Lambda$ as a function of time. The protocol
consists of the following steps:

i) Prepare an initial symmetry-broken state $\left|
\Psi(t_0)\right\rangle$ spread over many eigenstates in the degenerate
phase, with a certain initial value of the external parameter
$\Lambda_0$, and let it equilibrate to $\left|
\Psi(t_r)\right\rangle$. The equilibration time $t_r$ is long enough
to assure that all the relevant observables just fluctuate around the
equilibrium value.
 
ii) Perform the forward process from $\Lambda_0$ to $\Lambda_1$,
crossing the critical line and hence finishing in the normal phase,
linearly changing the external parameter. Then, let the system
equilibrate following a unitary evolution under $H(\Lambda_1)$.

iii) Perform the backward process from $\Lambda_1$ to $\Lambda_0$ to
reach the starting point, completing in this way the closed cycle, and
finally let it equilibrate under $H(\Lambda_0)$.

It is worth to remark that the equilibration time $t_r$ is required to
ensure that the system has equilibrated properly before any process,
forward or backward, is carried out. The velocity of the protocol is
defined as the rate of change of $\Lambda(t)$. For an enough slow
protocol the system remains always in equilibrium. However, this is not
the case for faster or sudden driving processes, for which the system
has no time to {\it feel} the changes in the Hamiltonian parameters
during the protocol, and therefore the whole equilibration occurs
after. Hence, we introduce the equilibration time $t_r$, during which
the system evolves under a fixed value of $\Lambda$, to ensure that,
even for faster rates of change, the system equilibrates before
starting the next step in the protocol.

Although we will make use of different initial states throughout this
article, coherent initial states play a very relevant role, because
they properly describe the ones with population imbalance engineered
in~\cite{Zibold:10}. They are given by $\left| \mu \right \rangle =
\mathcal{N} \sum_{m_j=-J}^{J}\mu^{m_j+J} \left| J,m_j\right \rangle $,
being $\mu\in\left[-1,1\right]$, $J_z \left|J,m_j
\right\rangle=m_j\left|J,m_j \right\rangle$ and $\mathcal{N}$ the
normalization constant, where a positive value of $\mu$ characterizes
a symmetry-breaking state in the positive branch $\langle J_x
\rangle>0$, and vice versa. We will rely on them to obtain the main
numerical results of this article, presented in Sec. III B. However,
in order to obtain a more complete picture and to stress the
generality of the reported irreversibility, in the Sec. III C we
consider different kinds of initial states: rectangular, Gaussian and
double-Gaussian.

The time evolution is dictated by the Schr\"{o}dinger equation
$i\frac{d\left| \Psi(t)\right\rangle}{dt}=H\left(\Lambda(t)\right)
\left| \Psi(t)\right\rangle$, which can be solved implementing the
method used in~\cite{Caneva:08}. In the basis $\left|J,m_j \right
\rangle$, the state at time $t$ formally reads $\left|\Psi(t) \right
\rangle = \sum_{m_j=-J}^J u_{m_j} \left(t \right) \left| J,m_j \right
\rangle$. Therefore, solving the following set of $2J+1$ coupled
equations
\begin{eqnarray}
\label{eq:time}
&i\frac{du_{m_j}(t)}{dt} =
\left(\frac{J(J+1)-m_j^2}{2}-\frac{J}{2\Lambda(t)}\right)u_{m_j}\left(t
\right) +& \\ \nonumber
&+\frac{\sqrt{J(J+1)-m_j(m_j+1)}\sqrt{J(J+1)-(m_j+1)(m_j+2)} }{4}
u_{m_j+2}\left(t \right)+&\\ \nonumber &
+\frac{\sqrt{J(J+1)-m_j(m_j-1)}\sqrt{J(J+1)-(m_j-1)(m_j-2)} }{4}
u_{m_j-2}\left(t \right),& 
\end{eqnarray}
one obtains the evolved state according to $\Lambda(t)$. We choose a
linear time dependence in $\Lambda$ that obeys the following function
\[   \Lambda(t)=\left\{
\begin{array}{ll}
\label{eq:lambda}
  \Lambda_0, & \text{if } t\in
  \left[0,t_r\right)\\ \Lambda_0+\Delta\Lambda \frac{t-t_r}{\tau_q}, &
    \text{if } t \in \left[t_r,t_r+\tau_q \right) \\ \Lambda_1, &
      \text{if } t\in \left[t_r+\tau_q,2t_r+\tau_q\right)
        \\ \Lambda_1-\Delta\Lambda \frac{t-(2t_r+\tau_q)}{\tau_q}, &
        \text{if } t \in \left[ 2t_r+\tau_q,2t_r+2\tau_q \right)
          \\ \Lambda_0, & \text{if } t\in
          \left[2t_r+2\tau_q,3t_r+2\tau_q\right].
\end{array} 
\right. \] 
where $\Delta\Lambda=\Lambda_1-\Lambda_0$, $t_r$ is the selected time
to ensure the equilibration, and $\tau_q$ the driving time (see inset
of Fig.~\ref{fig:energy} for more details). Note that the time
evolution is always unitary. The only energy exchange is the work done
by the protocol $\Lambda(t)$; neither a thermal bath, nor any other
kind of environment is coupled to the system at any time.

The rapidity of the driving is determined by a time scale $\tau_s$,
related to an effective gap, $\tau_s\sim 1/\Delta^{eff}$. Since the
state is driven across an ESQPT, the energy difference between states
with same parity, $\epsilon_i^{\pm}(\Lambda)=\left|
E_{i,\pm}(\Lambda)-E_{i-1,\pm}(\Lambda) \right|$, reaches a minimum
value, $\Delta^{\pm}_i=\min_{\Lambda}\epsilon_i^{\pm}(\Lambda)$ being
$\Lambda \in \left[ \Lambda_1,\Lambda_0 \right]$. We rely on this fact
to get a reasonable estimation of $\tau_s$, considering
$\Delta^{eff}=\min_{i,\pm} \Delta_i^{\pm}$.

\begin{figure}
  \includegraphics[width=0.7\linewidth,angle=-0]{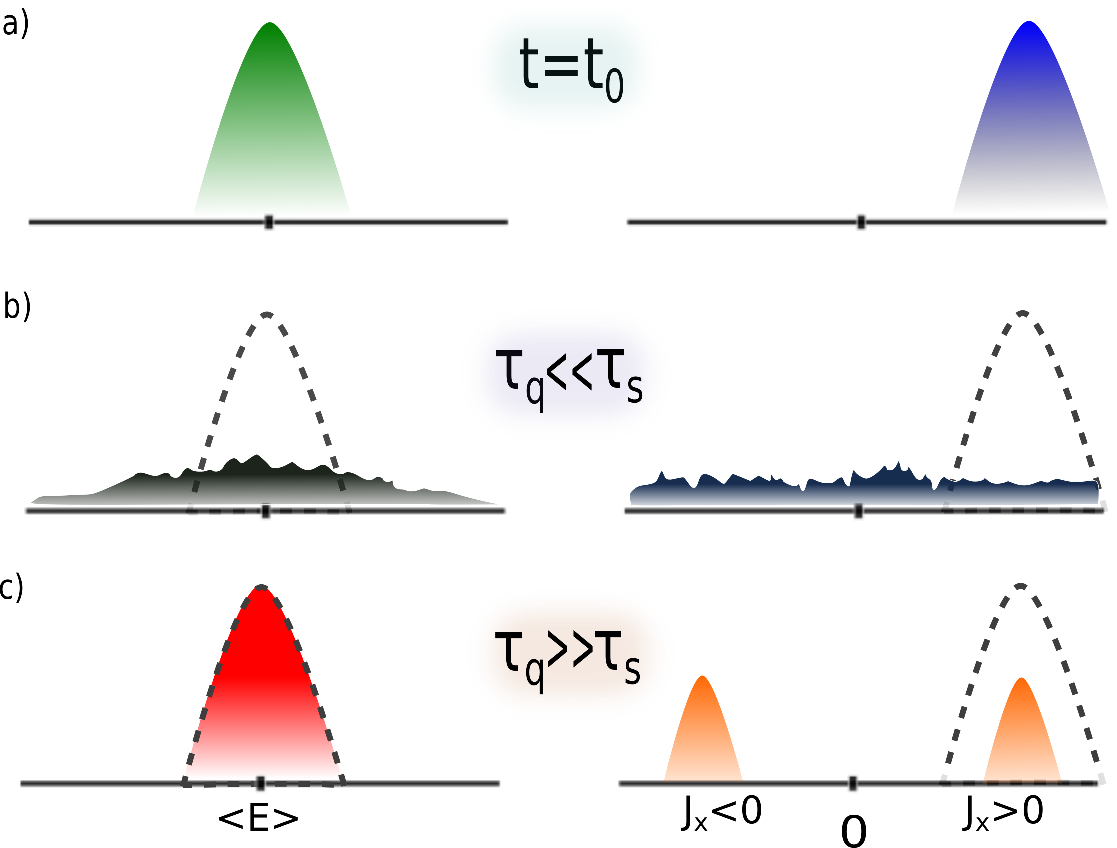}
  \caption{Scheme of the processes: probability distributions of the
    energy (left) and $J_x$ (right).  a) Symmetry-breaking initial
    state, with mean energy $\left \langle E\right\rangle$ and $ \left
    \langle J_x \right \rangle \neq 0$. b) A comparison between the
    initial distributions (dashed lines) and the final ones for a fast
    driving, and c) for a slow driving. If $\tau_q \ll\tau_s $, the
    energy is largely dissipated. If $\tau_q \gg \tau_s$ the energy
    distribution is fully recovered, but the process is irreversible
    since both modes are equally populated and therefore $\left<
    J_x\right>=0$.}
\label{fig:scheme}
\end{figure}

\section{Results}

\subsection{Main result}

The main result of the present article is sketched in
Fig.~\ref{fig:scheme}. It can be summarized as follows:

If the driving is fast ($\tau_q/\tau_s \ll 1$) energy is largely
dissipated and the final state is totally different from the initial
one; this constitutes a standard irreversible process. On the
contrary, if the driving is very slow ($\tau_q/\tau_s \gg 1$) the
energy distribution is fully recovered; in other words, the total
energy necessary to complete the cycle is zero, and all the mechanical
work invested in the forward part is exactly recovered in the
backwards. However, the distribution of the order parameter $J_x$ is
dramatically changed. In the final state, this distribution is
symmetric around zero, independently of the initial expectation value
$\left\langle J_{x,i} \right\rangle$; it consists of two smaller
copies of the initial one, centered at $J_x=\left\langle J_{x,i}
\right\rangle$ and $J_x=-\left\langle J_{x,i} \right\rangle$,
respectively. This entails the loss of the information about the
initial symmetry-breaking, and can be quantified by the corresponding
increase of the information entropy $I(J_x) \sim \log 2$. Therefore,
the process is irreversible despite the total absence of energy dissipation.

\subsection{Unitary evolution with a coherent initial state}

We explore the non-equilibrium dynamics of the system by means of a
numerical simulation involving $N=500$ particles, where the estimated
time scale results to be $\tau_s\simeq 0.01$. The presented results
were obtained choosing a coherent state $\left|\mu=1/2 \right \rangle$
as initial condition, and an equilibration time $t_r=9\times 10^ 4
\tau_s$.  The initial and final values of the external parameter are
$\Lambda_0=7/2$ and $\Lambda_1 = 1/2$, respectively. To analyze
different dynamical regimes as a function of the driving time, we
study the expectation value of the first component of the Schwinger
angular momentum $\left \langle J_x(t) \right \rangle$, and the
probability distributions over $E_{i,n}$ and $J_x$, given by $\left|
\left \langle E_i,\pi_n \right| \left. \Psi(t) \right \rangle
\right|^2=\left | C_{i,n}(t) \right| ^2$ and $\left| \left \langle
J,j_x \right|\Psi (t) \left. \right \rangle \right|^2=\left|
K_{j_x}(t)\right|^2$ respectively, where $\left| E_i,\pi_n \right>$ is
an eigenstate of the Hamiltonian $H$, and $\left| J,j_x \right>$ an
eigenstate of $J_x$, that is $J_x\left| J,j_x \right>=j_x \left| J,j_x
\right>$.

\begin{figure}
  \includegraphics[width=0.45\linewidth,angle=-90]{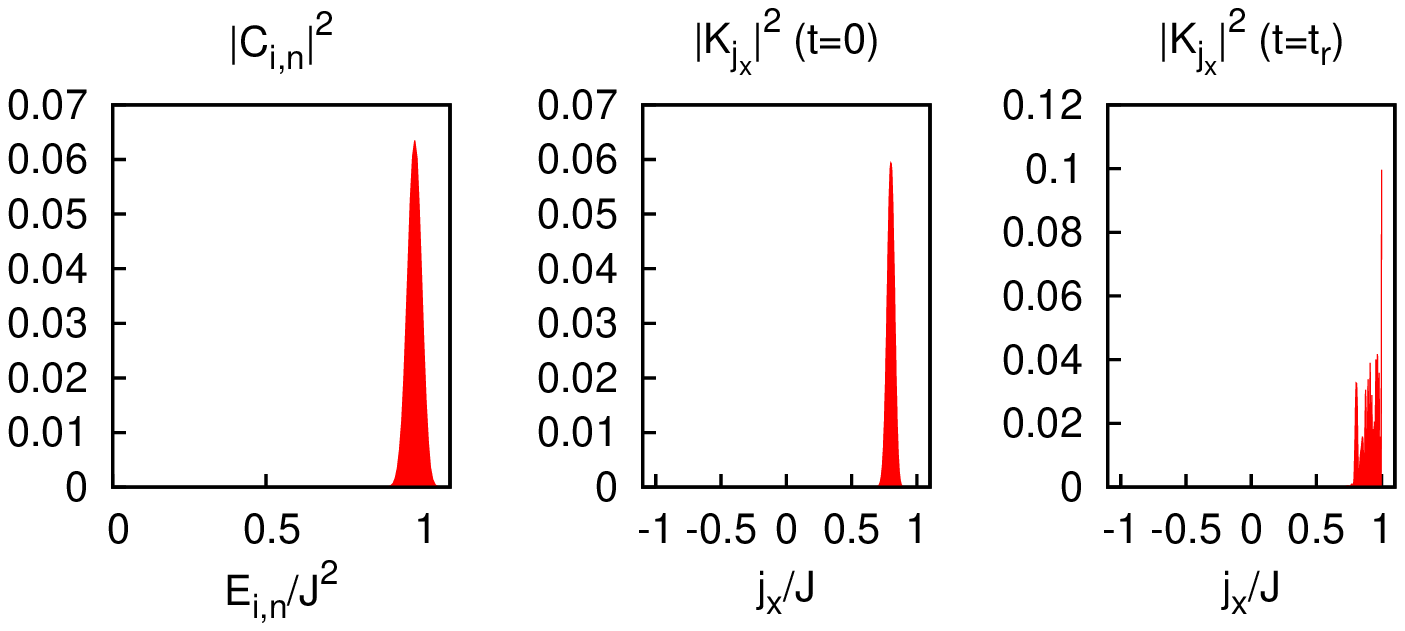}
  \caption{Probability distributions of the initial state. At the
    left, the energy distribution $\left |C_{i,n}(0) \right| ^2$ is
    displayed. In the middle and at the right, the probability
    distributions over $J_x$, $\left| K_{j_x}(t)\right|^2$, at $t=0$
    and $t=t_r$ are showed.}
  \label{fig:coherent}
\end{figure}

The initial distributions corresponding to the initial state are
represented in the Fig. \ref{fig:coherent}. We plot the energy
distribution before the protocol on the left, $\left | C_{i,n}(0)
\right| ^2$ (trivially, it remains unchanged between $t=0$ and
$t=t_r$); the initial distribution of $J_x$, $\left|
K_{j_x}(0)\right|^2$, in the middle, and the same distribution at
$t=t_r$, after the system is equilibrated, on the right.

\begin{figure}
  \includegraphics[width=0.72\linewidth,angle=-90]{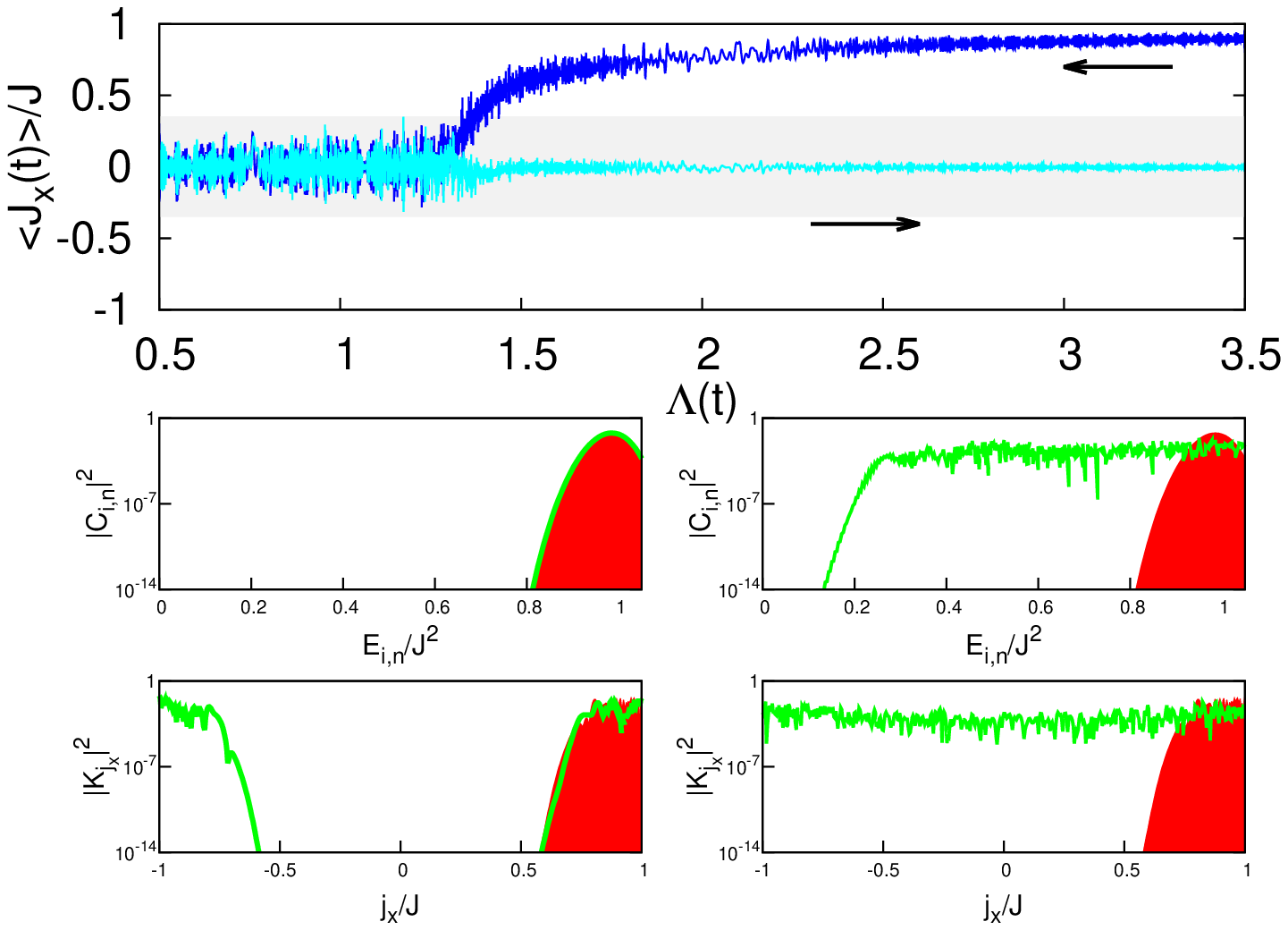}
  \caption{Results for two different dynamical regimes, left panels
    correspond to $\tau_q/\tau_s=7200$ while right panels correspond
    to $\tau_q/\tau_s=0.4$.  At the top, time evolution of the
    expectation value $\left \langle J_x(t) \right \rangle$ as a
    function of the value of the external parameter
    $\Lambda(t)$. Forward and backward processes are plotted with dark
    and light lines (dark blue and cyan online) respectively.  The
    arrows indicate the direction of the process starting at
    $\Lambda_0=7/2$.  The gray band represents the possible final
    results for the slow protocol (see main text for more details).
    Middle and bottom rows, probability distribution of the energy and
    $J_x$. Filled region (red online) represents the initial
    probability distribution at $t=t_r$; solid lines (green online)
    the final at $t=2\tau_q+3t_r$.}
\label{fig:cycle}
\end{figure}

In Fig.~\ref{fig:cycle} we summarize our main results. In the upper
part, the plot shows the time evolution of $\left\langle J_x(t)
\right\rangle$ for the slow driving ($\tau_q/\tau_s=7200$) ---the
trajectory $\left \langle J_x(t) \right \rangle$ does not provide
significative information for the fast driving ($\tau_q/\tau_s=0.4$),
since the system is not equilibrated at intermediate times. In the
panels below, we plot the main probability distributions, for the two
paradigmatic regimes: at the left, for a slow driving
($\tau_q/\tau_s=7200$); at the right for a fast driving
($\tau_q/\tau_s=0.4$).  In the middle panel, the energy distributions
$\left|C_{i,n}(t) \right|^2$ are shown before (filled region) and
after (solid lines) the cycle, in a logarithmic scale. In the lower
panel, the probabilities $\left|K_{j_x}(t) \right|^2$ are plotted in
the same format. It is clearly shown that the initial energy
distribution is fully recovered at the end of the slow cycle.  On the
contrary, $\langle J_x (t) \rangle$ returns following a path totally
different from the corresponding to the forward protocol. During the
backward process, $\langle J_x(t) \rangle$ always fluctuates around a
value close to zero, implying that the state consists of a
superposition of the two symmetry-breaking branches, as it can be
observed in the lower panel for $\left|K_{j_x}(t) \right|^2$.  To get
a deeper insight on this fact, we take into account that the
equilibrium properties of the final state can be described by means of
a long-time average at the fixed value $\Lambda_0$, from which we
obtain the final equilibrium value
$\overline{\left<J_{x,f}\right>}$. It is important to note that the
returning path does not average exactly to $\overline{\left\langle
  J_{x,f} \right\rangle}=0$, but to a certain finite value depending
on both the driving and the equilibration times, $\tau_q$ and
$t_r$. We have performed several calculations with different
equilibration times, which result in a region of possible final values
of $\overline{\left<J_{x,f} \right>}$ at $\Lambda_0$, plotted as a
band in the upper part of Fig. \ref{fig:cycle}.  Therefore, the final
distribution of $J_x$ is not exactly symmetric in the majority of the
cases and the superposition of the two branches is slightly biased
towards the left or the right. Hence, for finite-size systems not all
the information about the initial symmetry-breaking, but a very
significant part of it, is lost as a consequence of the phase mixing
between eigenstates of opposite parity once the state enters in the
region without degeneracies.  In Sec. III C we will show how the width
of the band decreases as the number of initially populated eigenstates
increases, suggesting that the band tends to shrink to
$\overline{\left\langle J_x \right\rangle}=0$ in the thermodynamic
limit. Furthermore, in Sec. V we will give theoretical arguments
supporting this conclusion. Although not explicitly shown, the same
final state is obtained independently of the degree of
symmetry-breaking of the initial condition.

On the other hand, the fast protocol corresponds to a standard
irreversible process, with measurable consequences in any
observable. In particular, the final distribution for the energy is
totally different from the initial one. A thermodynamic interpretation
of this fact can be done in the following terms. The energy variation
in a process done in a quantum isolated system $\Lambda_i \rightarrow
\Lambda_f$ can be divided in two parts: $\Delta E = W + Q$, with
$W=\sum_j p_j^i \left[ E_j \left( \Lambda_f \right) - E_j \left(
  \Lambda_i \right) \right]$, and $Q= \sum_j \left[ p_j^f - p_j^i
  \right] E_j \left( \Lambda_f \right)$, being $E_j (\Lambda)$ the
j-th energy level of $H(\Lambda)$, and $p_j^i$ and $p_j^f$ the
population of that energy level in the initial (before the protocol)
and in the final state (after completing the protocol),
respectively. $W$ can be understood as the reversible work done in the
system, and $Q$ as the dissipated heat as a consequence of the
protocol~\cite{Polkovnikov:08}. $W$ represents the energy change due
to the change in the energy levels $E_j (\Lambda_i) \rightarrow E_j
(\Lambda_f)$, which only depends on the initial and final values of
the external parameter of the Hamiltonian. On the other hand, $Q$
represents the change in the energy due of the non-adiabatic
transitions between energy levels; this is the reason why it can be
understood as the heat dissipated by the protocol, despite the system
is always isolated from any external environment. Therefore, if the
energy distribution remains unchanged after a cyclic process, that is,
the population of all the energy levels is the same in the initial and
the final states, $p_j^f = p_j^i$ $\forall j$, the mechanical work
invested in the forward process is exactly retrieved in the backward,
and therefore the process is usually understood as reversible. On the
contrary, any change in the energy distribution after a cyclic process
implies that some of this mechanical work has been dissipated into
heat, meaning that the population of the energy levels has changed
$p_j^f \ne p_j^i$ and hence the process is irreversible (see Sec. IV
for a link between this dissipated energy and the von Neumann entropy
of the equilibrium states). Furthermore, in this last case (fast
driving protocol) as we can see in Fig.~\ref{fig:cycle} (or in
Fig.~\ref{fig:cycle_additional} for additional initial states), the
final distribution of $J_x$ is also totally different from the initial
one. First, it is symmetric, implying that $\left< Jx\right>=0$, as it
also happens as a consequence of the slow driving. Second, it is much
wider, with a shape totally different from the initial one; it does
not consist of the superposition of two modes, but to a roughly flat
distribution between $J_x=-J$ and $J_x=J$, which is a consequence of
the energy dissipation of the process.

\subsection{Additional initial states}

We provide additional results with different initial symmetry-breaking
states.  The mean spacing of the spectrum at $\Lambda_0=7/2$ is
$\bar{E}/J^2\simeq 0.003$, which turns out to be an useful quantity to
compare the width of the states. Here, we define three different
initial states to support the previous conclusions based on the
driving process starting with a coherent state $\left|\mu=1/2 \right
\rangle$. In all of them only one mode is populated and therefore the
population imbalance is maximum with $\left \langle J_x \right \rangle
\simeq J$:
\begin{itemize}
\item {\em Rectangular}: An initial state that just populates the last
  $50$ double-degenerated eigenstates ($M=25$ doublets) of the
  Hamiltonian with the same probability, i.e. $\left|C_{i,n}
  \right|_R^2=1/(2M)$ for the last $2M=50$ eigenstates
  $\left|E_i,\pi_n \right\rangle$, and $\left|C_{i,n} \right|_R^2=0$
  for the rest.
\item {\em Gaussian}: An initial state that populates the eigenstates
  with a Gaussian probability, whose mean energy value is
  $\tilde{\mu}_{1G}/J^2=\left \langle E_{1G}\right \rangle/J^2=0.922$,
  and variance $\sigma_{1G}/J^2=0.035 \simeq 10
  \bar{E}/J^2$. Therefore, an eigenstate $\left|E_i,\pi_n \right
  \rangle$ will be populated according to $\left|C_{i,n}
  \right|_{1G}^2\propto
  e^{-\frac{\left(E_{i,n}-\mu_{1G}\right)^2}{2\sigma_{1G}^2}}$.
  
\item {\em Double-Gaussian}: An initial state that consists of two
  separated Gaussian probability distributions. Both have the same
  variance $\sigma_{2G}/J^2=0.017\simeq 5 \bar{E}/J^2$ but different
  mean energy, which are $\tilde{\mu}_{2G_1}/J^2=0.971$ and
  $\tilde{\mu}_{2G_2}/J^2=0.805$. Therefore $\left\langle E_{2G}
  \right \rangle/J^2 \simeq 1/2
  \left(\tilde{\mu}_{2G_1}+\tilde{\mu}_{2G_2}\right)=0.888$. Hence, an
  eigenstate $\left|E_i,\pi_n \right \rangle$ will be populated
  according to $\left|C_{i,n} \right|_{2G}^2\propto
  e^{-\frac{\left(E_{i,n}-\mu_{2G_1}\right)^2}{2\sigma_{2G}^2}}+e^{-\frac{\left(E_{i,n}-\mu_{2G_2}\right)^2}{2\sigma_{2G}^2}}$.
\end{itemize}

\begin{figure}
  \centering
  \includegraphics[width=0.72\linewidth,angle=-90]{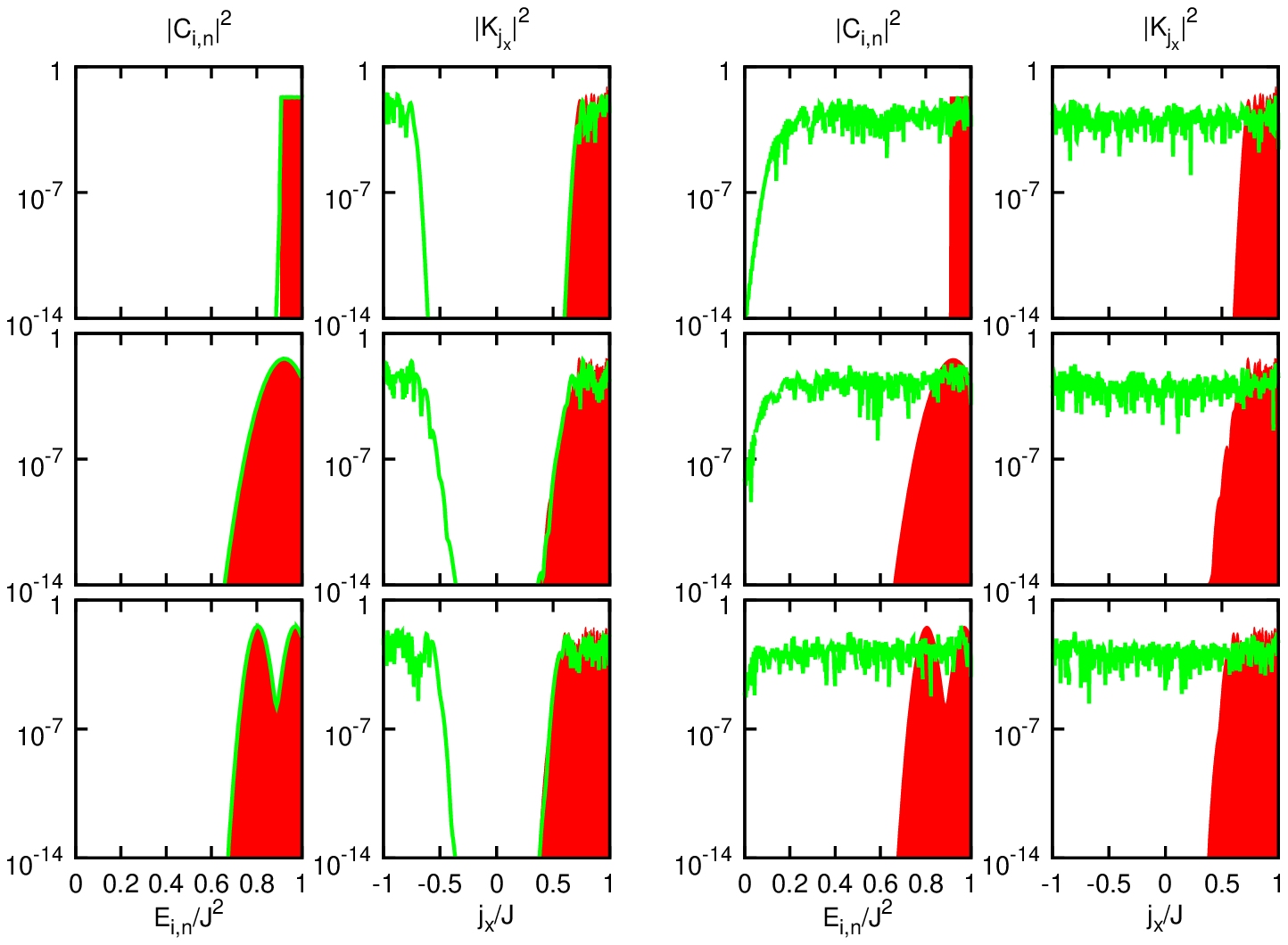}
  \caption{Results for two different dynamical regimes for additional
    initial states, the two columns at the left correspond to a slow
    driving ($\tau_q/\tau_s=1538$), whereas the two columns at the
    right correspond to a fast driving ($\tau_q/\tau_s=0.2$). The
    results for rectangular, Gaussian and double-Gaussian are plotted
    in the first, second and third row respectively. For each driving
    process, we represent the probability distribution of the energy
    (left) and of the $J_x$ (right) in logarithmic scale. The filled
    region (red online) represents the initial probability
    distribution at $t=t_r$, while solid lines (green online) the
    final at $t=2\tau_q+3t_r$.}
\label{fig:cycle_additional}

\end{figure}

After performing a protocol for each additional initial state, we
obtain the results summarized in the different panels of the
Fig.~\ref{fig:cycle_additional}. Each row corresponds to a different
initial state, rectangular (top), Gaussian (middle) and
double-Gaussian (bottom). For each case, a fast ($\tau_q/\tau_s=0.2$)
and a slow ($\tau_q/\tau_s=1538$) protocol are presented. As we have
already showed in the previous section, while for the slow protocol
the energy distribution is completely recovered and the probability
distribution over $J_x$ is split in the two possible branches ($\left<
J_x\right> \lessgtr 0$), for a fast protocol neither the
$J_x$-distribution nor the energy are recovered, which implies large
dissipation.

\begin{figure}
  \centering
  \includegraphics[width=0.70\linewidth,angle=-90]{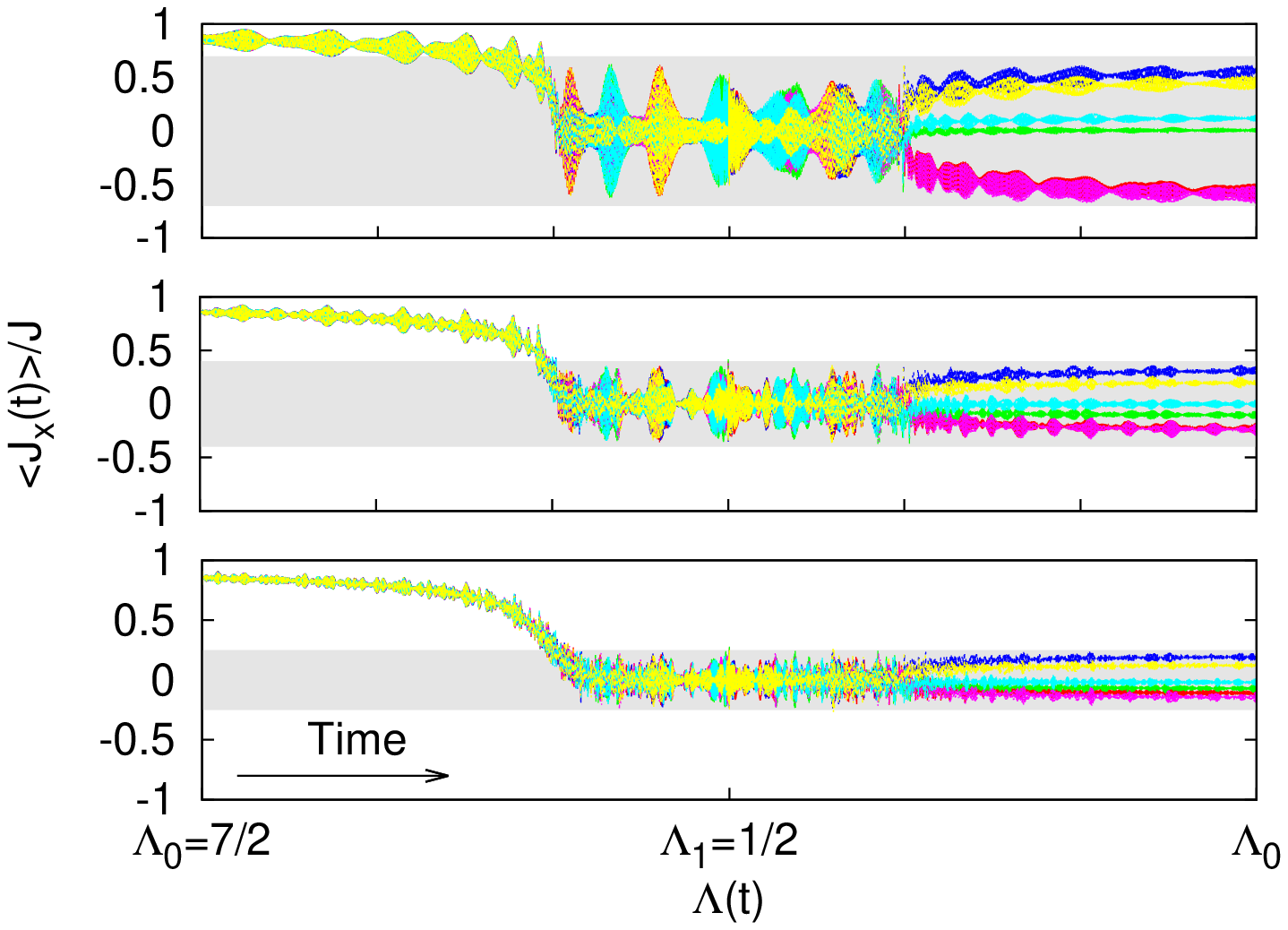}
  \caption{Expectation value $\left< J_x(t) \right>$ as a function of
    the time-dependent control parameter $\Lambda(t)$ for Gaussian
    initial states with different variances: $\sigma \simeq 2.3
    \bar{E}$ (top), $\sigma \simeq 7 \bar{E}$ (middle) and $\sigma
    \simeq 16.3\bar{E}$ (bottom), but same mean energy. For each
    Gaussian initial state, we perform six cycles with slightly
    different values of $\tau_q$ times (from $\tau_q/\tau_s=1500$ to
    $\tau_q/\tau_s=1505$) plotted with different colors. For all the
    cases, the initial value $\left<J_x (t=0) \right> \simeq J$,
    whereas the possible final results are encoded by a gray
    band. Clearly, the wider the initial state, the narrower the gray
    band.}
\label{fig:band}
\end{figure}

Finally, we perform different cycles for three Gaussian initial states
with the same mean energy but different variances, in order to study
how the width of the band shown in the upper part of
Fig.~\ref{fig:cycle} depends on the number of initially populated
double-degenerated eigenstates. We consider three Gaussians with
variances $\sigma \simeq 2.3 \bar{E}, 7 \bar{E}$ and $16.3\bar{E}$,
respectively. Then, we perform slow cycles with slightly different
values of $\tau_q$ to obtain the possible final results of the
expectation value of the observable $J_x$, ensuring that the energy
distribution is recovered. The results are summarized in the
Fig.~\ref{fig:band}. There, the narrowest Gaussian initial state (top)
generates a broad region of possible final expectation values $\left<
J_x\right>$, which means that the protocol can end in a state very
similar to the initial condition. On the other hand, the widest
Gaussian initial state (bottom) provides a narrow band, since the
number of initially populated eigenstates is considerably larger. This
entails that the protocol always ends with an almost perfect
superposition of the two symmetry-breaking branches. Clearly, the
wider the initial state, the narrower the gray band of the possible
final results, and hence is reasonable to conjecture that the band
shrinks to zero, and then, the final expectation value $\left<
J_x\right> \rightarrow 0$ in the thermodynamic limit.  This fact will
be confirmed with theoretical arguments in the Sec. V.

\section{Entropy and information} 

As the state of the system is always pure, we cannot link this source
of irreversibility to a (thermodynamic) entropy production. To develop
a thermodynamic interpretation, we rely on initial (before the
protocol) and intermediate (before the backward process) equilibrium
states, which are the key elements of the Crook's
theorem~\cite{Talkner:08,Crooks:98}. It has been recently shown that
these equilibrium states exist for (almost) any isolated quantum
system unitary evolving after (almost) any initial condition, and they
coincide with the long-time average, $\rho_{\text{eq}}=\lim_{T
  \rightarrow \infty} \frac{1}{T} \int_0^T dt \, \left| \Psi(t)
\right> \left< \Psi(t) \right|$~\cite{Reimann:12} ---the actual state
$\left| \Psi(t) \right\rangle$ fluctuates around this equilibrium
state, remaining close to it during the majority of the time.  Note
that this is just a formal definition, meaning a very long time
average keeping fixed all the free parameters of the Hamiltonian. In
our case, it has to be understood as a time average of the state
evolving under $H(\Lambda)$, with a fixed value of $\Lambda$, not to
an average over the whole protocol $\Lambda(t)$.

\begin{figure}
  \includegraphics[width=0.4\linewidth,angle=-90]{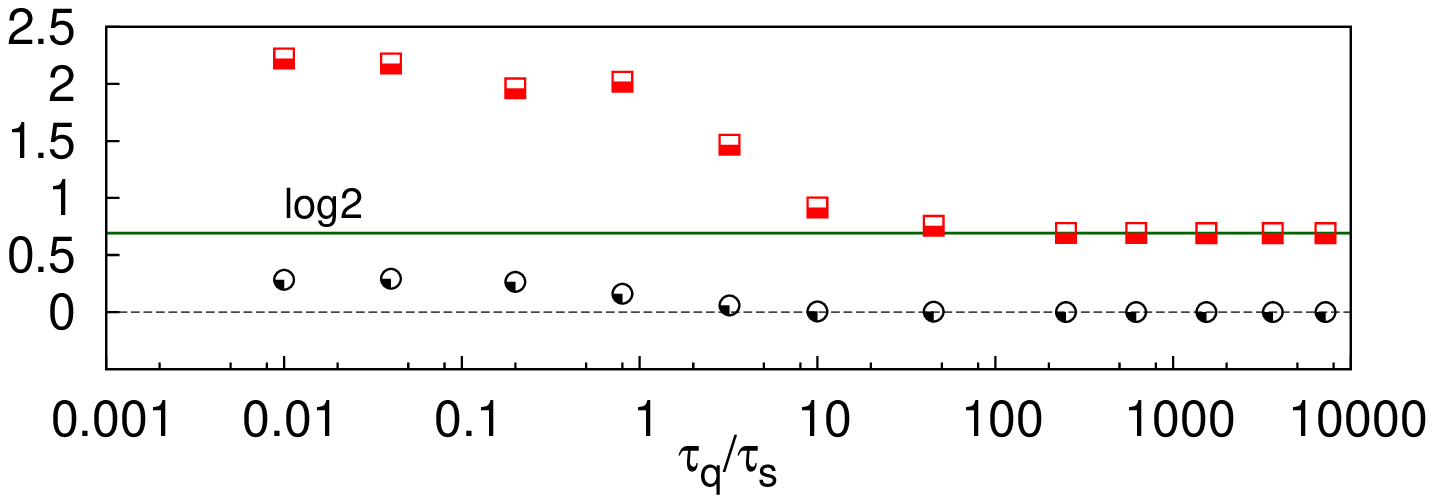}
  \caption{Increment of the von Neumann entropy $\Delta S$ between the
    initial state and the equilibrium state after the forward process
    with squares (red online), as a function of the driving time
    $\tau_q/\tau_s$. The circles (black online) illustrate the
    dissipated energy $\left| \left\langle E_{\text{dis}}
    \right\rangle \right|/\left \langle E _{\text{initial}}\right
    \rangle$.}
\label{fig:entropy}
\end{figure}

From this equilibrium state, we calculate $\Delta S = S(\Lambda_1) -
S(\Lambda_0)$, being $S = - \text{Tr} \left[ \rho_{\text{eq}} \log
  \rho_{\text{eq}} \right]$ the von Neumann entropy of these reference
states. We compare this value with the dissipated energy, $
\left\langle E_{\text{dis}}\right\rangle = \left\langle
E_{\text{final}}\right\rangle - \left\langle
E_{\text{initial}}\right\rangle$. Results are plotted in
Fig.~\ref{fig:entropy}, where the circles represent the dissipated
energy $\left| \left\langle E_{\text{dis}} \right\rangle
\right|/\left\langle E_{\text{initial}}\right\rangle$ and the squares
the increment of entropy $\Delta S$, as a function of the driving time
$\tau_q/\tau_s$. When this time is short, $\tau_q\ll\tau_s$, the
energy is largely dissipated, due to the non-adiabatic transitions
between energy levels, as it is discussed in Sec. III; hence, the
increment of entropy is substantial. The number of these non-adiabatic
transitions diminishes with the increase of the driving time $\tau_q$,
decreasing the von Neumann entropy in a similar rate. Finally, if
$\tau_q\gg\tau_s$, there is no dissipated energy, but the final
entropy is still larger than the initial one, $\Delta S = \log 2$.

This result contradicts the common thermodynamic lore, based on a
direct link between entropy production and energy dissipation. From
Figs.~\ref{fig:cycle} and~\ref{fig:entropy} we infer that, for slow
enough processes $\tau_q \gg \tau_s$, the probability of investing a
certain work $w$ in the forward process, $P_f(w)$, is exactly the same
than the probability of recovering the same magnitude of work $-w$ in
the backwards $P_b(w)$, that is, $P_f(w)/P_b(-w)=1$. This implies that
any measurement of the dissipated work $w_{\text{dis}}$ in the
complete protocol, which can experimentally performed following the
strategy very recently proposed in~\cite{Roncaglia:14}, gives
$w_{\text{dis}}=0$; the population of all energy levels remains
unchanged, no heat is produced and all the work done in the forward
part of the protocol is retrieved in the backwards. On the contrary, a
similar measurement performed on the observable $J_x$, which can be
experimentally done following~\cite{Zibold:10}, gives an opposite
result: the probability of obtaining the same value $j_x$ is not the
same at the beginning and at the end of the protocol, $P_f (j_x) \ne
P_i (j_x)$.

To obtain a physical interpretation of this fact, we consider the
information entropy of an observable $A$, defined as $I\left( A\right)
= - \sum_n p_n \log p_n$, being $p_n$ the probability of obtaining the
eigenvalue $A_n$ in a measurement. From the previous results it is
clear that $\Delta I(H) \rightarrow 0$ when $\tau_q \gg \tau_s$. On
the contrary, $\Delta I(J_x) \sim \log 2$~\footnote{This is not an
  exact result because $J_x$ is not constant in time, but it
  fluctuates around the equilibrium value, so its probability
  distribution also fluctuates with time.} when $\tau_q \gg \tau_s$.
Hence, the main consequence of the protocol is a loss of information
about $J_x$ which equals the increase of the von Neumann entropy of
the long time-average state, despite the absence of dissipation. This
constitutes an unexpected source of irreversibility.

A qualitative explanation of this fact can be done in the following
terms. The direct link between entropy production and energy
dissipation is established considering that the final equilibrium
value after any thermodynamic process, $\rho_{\text{eq}}$, is given by
a density matrix which is {\em diagonal} in the basis that
diagonalizes the CSCO~\cite{Polkovnikov:11b}. The precise shape of
this matrix is determined by all the relevant conserved quantities of
the Hamiltonian~\cite{Polkovnikov:11}. In our case, the only global
constant of motion for any value of the coupling parameter $\lambda$
is the parity $\Pi$. Hence, at a first sight, the equilibrium state
$\rho_{\text{eq}}$ should be diagonal in the eigenbasis which
diagonalizes the $\{H, \Pi\}$, that we label $\left| E_i, \pm
\right>$. However, there exists another operator which remains
constant only above the critical energy of the ESQPT, the coherences
between the two parity sectors of the Hamiltonian $C=\sum_i \left|
E_i, + \right> \left< E_i, - \right| + h. c.$ The key point is that
this operator has no diagonal elements in the eigenbasis of $\{ H, \Pi
\}$, and hence any information regarding its initial value is encoded
in the non-diagonal part of the initial equilibrium state. This
information is kept by the protocol only if the system remains above
the critical energy of the ESQPT; as soon as the protocol leads the
system to $E<E_c$, $C$ ceases to be a constant of motion and all the
information about its initial value is erased by phase mixing. As a
consequence, $\rho_{\text{eq}}$ changes from non-diagonal in the
initial state, to diagonal in the final equilibrium state, leading
$\Delta S = \log 2$. And, as we have pointed before, this irreversible
change persists even when the protocol is performed slow enough to
avoid non-adiabatic transitions between energy subspaces, and
therefore constitutes a source of irreversibility independent of the
energy dissipation.

\section{Mechanism of irreversibility} 

The microscopic origin of this kind of irreversibility is the
collective phase mixing in the normal phase.  The effect of the
intrinsic irreversibility due to the loss of information is remarkable
when the process is performed slow enough to prevent transitions
between different energy levels, that is, when the system evolves in
the quasistatic limit.  Hence, the wave function can be expressed as
$\left| \Psi(t)\right \rangle=\sum_{i,n}e^{-i \phi_{i,n}
  (t)}\Gamma_{i,n} \left| E_i,\pi_n \right \rangle$, where
$\Gamma_{i,n}$ are the coefficients of the initial wavefunction in the
eigenstates of $H\left(\Lambda_0\right)$, and $\left| E_i,\pi_n \right
\rangle$ are the common eigenstates of the instantaneous Hamiltonian
$H\left(\Lambda(t) \right)$ and the parity $\Pi$. Therefore, the only
relevant change in the wave-function due to the time-dependent
protocol $\Lambda(t)$ resides in the phases $\phi_{i,n} = \theta_{i,n} +
\gamma_{i,n}$. The first term $\theta_{i,n}$ accounts for the
dynamical phase, $\theta_{i,n} (t) = \int_0^t d \tau E_{i,n}
(\tau)$. The second one represents the geometrical Berry phase,
$\gamma_{i,n} = -i \int_{\Lambda_i}^{\Lambda_f} \left\langle
E_{i},\pi_n (\Lambda) \right| \partial_{\Lambda} \left| E_{i},\pi_n
(\Lambda) \right\rangle d \Lambda$, where the dependence on $\Lambda$
of the instantaneous eigenstates $\left|E_i,\pi_n \right>$ is
explicitly shown.  As our protocol consists in just changing one
parameter with the same initial and final states $\Lambda_i
\rightarrow \Lambda_f \rightarrow \Lambda_i$, the net contribution of
the Berry phase is zero, and hence the only relevant term is the
dynamical phase $\phi_{i,n} = \theta_{i,n}$. At the end of the cycle,
the value of this phase is
\begin{eqnarray}
\phi_{i,n} = \frac{2\tau_q}{\Lambda_1 - \Lambda_0}
\int_{\Lambda_0}^{\Lambda_1} &d \Lambda& E_{i,n}( \Lambda)+ \nonumber
\\ &+&t_r \left[2 E_{i,n}\left(\Lambda_0
  \right)+E_{i,n}\left(\Lambda_1 \right) \right].
\end{eqnarray}
To evaluate the changes in the wavefunction due to the protocol
$\Lambda_0 \rightarrow \Lambda_1 \rightarrow \Lambda_0$, we study how
the initial and final states evolve under $H(\Lambda_0)$. In this
case, the initial and the final time-dependent wavefunctions are
\begin{eqnarray}
\left| \Psi_i (t) \right> &=& \sum_{j,n} \text{e}^{-i E_{j,n} t}
\Gamma_{j,n} \left|E_j, \pi_n \right>; \\ \left| \Psi_f (t) \right>
&=& \sum_{j,n} \text{e}^{-i \phi_{j,n}} \text{e}^{-i E_{j,n} t}
\Gamma_{j,n} \left|E_j, \pi_n \right>.
\end{eqnarray}
So, the initial and final expectation values of any observable $O$
measured under such circumstances are
\begin{eqnarray}
\left< O_i (t) \right> = \sum_{kj} \sum_{nm} &&\text{e}^{-i (E_{k,n} -
  E_{j,m}) t} \times \nonumber \\ &\times&\Gamma^*_{j,m} \Gamma_{k,n}
\left< E_j, \pi_m \right| O \left| E_k, \pi_n \right>; \\ \left< O_f
(t) \right> = \sum_{kj} \sum_{nm}&& \text{e}^{-i (\phi_{k,n} -
  \phi_{j,m})} \text{e}^{-i (E_{k,n} - E_{j,m}) t}\times \nonumber
\\ &\times&\Gamma^*_{j,m} \Gamma_{k,n} \left< E_j, \pi_m \right| O
\left| E_k, \pi_n \right>. 
\end{eqnarray}

These expressions give the exact time evolution of $\langle O \rangle$
before and after completing the protocol. As we have previously shown,
these time-dependent expectation values fluctuate around the
corresponding equilibrium values, $\overline{\left\langle O_i
  \right\rangle}$ and $\overline{\left\langle O_f \right\rangle}$,
given in both cases by the long-time average under the same
Hamiltonian $H(\Lambda_0)$, $\overline{\left\langle O \right\rangle} =
\lim_{T \rightarrow \infty} \frac{1}{T} \int_0^T dt \left< O(t)
\right>$. Hence, to obtain a simple interpretation of the results, we
follow the reasoning in terms of these equilibrium values, which are
given by
\begin{eqnarray}
\overline{\left\langle O_i \right\rangle} &=& \sum_{kj} \sum_{nm}  \Gamma^*_{j,m} \Gamma_{k,n}
\left< E_j, \pi_m \right| O \left| E_k, \pi_n \right> \times \nonumber
\\  &\times&\lim_{T \rightarrow \infty} \frac{1}{T} \int_0^T dt
\text{e}^{-i (E_{k,n} - E_{j,m}) t}; \\ \overline{\left\langle O_f \right\rangle} &=& \sum_{kj}
\sum_{nm} \text{e}^{-i (\phi_{k,n} - \phi_{j,m})} \Gamma^*_{j,m}
\Gamma_{k,n} \left< E_j, \pi_m \right| O \left| E_k, \pi_n \right>
\times \nonumber \\ &\times& \lim_{T \rightarrow \infty} \frac{1}{T}
\int_0^T dt \text{e}^{-i (E_{k,n} - E_{j,m}) t}. 
\end{eqnarray}

Since $\Lambda_0$ lies on the degenerate phase, we can consider that
$E_{i,+} = E_{i,-}$, where $+$ ($-$) denotes the positive (negative)
parity sector. So, after the time average is performed, only diagonal
and non-diagonal terms connecting $\left|E_i, + \right>$ with
$\left|E_i, - \right>$ survive. Therefore, the initial and final
equilibrium values are
\begin{eqnarray}
\overline{\left\langle O_i \right\rangle} &=& \sum_j \left| \Gamma_{j,+} \right|^2 \left< E_j,
+\right| O \left| E_j, + \right> + \nonumber \\ &+& \sum_j \left|
\Gamma_{j,-} \right|^2 \left<E_j, - \right| O \left| E_j, - \right> +
\\ &+& \sum_j \left( \Gamma_{j,+}^* \Gamma_{j,-} \left<E_j, + \right|
O \left| E_j, - \right> + h. c. \right); \nonumber  \\ \overline{\left\langle O_f \right\rangle}
&=& \sum_j \left| \Gamma_{j,+} \right|^2 \left<E_j, + \right| O \left|
E_j, + \right> + \nonumber \\ &+& \sum_j \left| \Gamma_{j,-} \right|^2
\left<E_j, - \right|  O \left| E_j, - \right> + \\ &+&\sum_j \left(
\Gamma_{j,+}^* \Gamma_{j,-} \text{e}^{-i \left( \phi_{j,+} -
  \phi_{j,-} \right)} \left<E_j, + \right| O \left| E_j, - \right> +
h. c. \right). \nonumber 
\end{eqnarray}
The phases affecting the non-diagonal part of the time average can be
written
\begin{eqnarray}
\phi_{j,+} - \phi_{j,-} &=& \frac{2 \tau_q}{\Lambda_1 - \Lambda_0}
\int_{\Lambda_0}^{\Lambda_1} d \Lambda \left[ E_{j,+} (\Lambda) -
  E_{j,-}(\Lambda) \right]+\nonumber
\\ &+&t_r\left[E_{j,+}\left(\Lambda_1 \right)- E_{j-}\left(\Lambda_1
  \right)\right] \equiv \omega_j.
\end{eqnarray}
As it is obvious from this equation, the phases $\omega_j = \omega_j
(\tau_q, t_r)$ depend on both the driving time, $\tau_q$, and the
equilibration time, $t_r$, though this dependence is not explicitly
written to simplify the notation.

A logical consequence of this fact is that the expected value of any
observable commuting with $\Pi$ is the same at the initial and the
final stages of the protocol. In particular, this happens for the
projector over any eigenstate, and hence the energy distribution
remains unchanged by the protocol. On the other hand, the result for
$J_x$ is
\begin{eqnarray}
\overline{\left\langle J_{x,i} \right\rangle} &=& \sum_j
\left(e^{-i\beta_j} \left|\Gamma_{j,\pm} \right| \left<E_j, + \right|
J_x \left| E_j, - \right> + h. c. \right) \\ \overline{\left\langle
  J_{x,f} \right\rangle} &=& \sum_j \left( \text{e}^{-i
  (\omega_j+\beta_j)} \left|\Gamma_{j,\pm} \right| \left<E_j, +
\right| J_x \left| E_j, - \right> + h. c. \right), \nonumber
\end{eqnarray}
where $\left|\Gamma_{j,\pm}
\right|e^{-i\beta_j}=\Gamma_{j,+}^*\Gamma_{,j-}$.  Hence, for slow
enough processes, the only consequence at the end of the cycle is a
set of phases $\omega_j$ arising for each doublet, giving rise to
$\overline{\left\langle J_{x,f} \right\rangle} = 2 \sum_j
\cos(\omega_j+\beta_j) \left|\Gamma_{j,\pm} \right|\left< E_j,+
\right| J_x \left| E_j, - \right>$. The exact value of this sum
depends on the initial condition, the details of the protocol and the
expected values $\left< E_j,+ \right| J_x \left| E_j, - \right>$, so a
numerical diagonalization of the quantum problem is required. However,
a good estimate can be obtained under certain assumptions. First,
consider that the initial condition has only real coefficients,
$\beta_j= n \pi, n = 0, 1, 2, \ldots, \, \forall j$, and suppose that
a mechanism similar to the Eigenstate Thermalization
Hypothesis~\cite{Srednicki:94,Rigol:08} holds, implying $\left<E_{j,+}
\right| J_x \left| E_{j,-} \right> \sim J_x^{\text{ETH}}$ for all the
$M$ populated eigenspaces; this entails that
\begin{equation}
\overline{\left\langle J_{x,f} \right\rangle} = 2 J_x^{\text{ETH}}
\sum_j^M \cos \omega_j \cos \beta_j \left|\Gamma_{j,\pm}\right|.
\end{equation}
Second, let's consider that the dynamical phases $\omega_j$ and the
coefficients $\cos \beta_j \left|\Gamma_{j,\pm} \right|$ behave as
uncorrelated random variables~\footnote{This is reasonable because the
  phases depend on the protocol $\Lambda(t)$, as a function of the
  trajectories $E_{j,n}(t)$, and thus are the same for any initial
  condition evolving under the same protocol.}. For two uncorrelated
random variables $X$ and $Y$, $\langle XY \rangle = \langle X \rangle
\langle Y \rangle$, being $\langle \rangle$ the statistical average,
and thus $\sum_{j=1}^M X_j Y_j = M \langle XY \rangle = M \langle X
\rangle \langle Y \rangle = (1/M) \sum_{j=1}^M X_j \sum_{j=1}^M
Y_j$. Applying this result we obtain
\begin{equation}
\overline{\left\langle J_{x,f} \right\rangle} = \frac{2 J_x^{\text{ETH}}}{M} \sum_j^M \cos \beta_j \left|\Gamma_{j,\pm}\right| \sum_j^M \cos \omega_j.
\end{equation}
Finally, considering that the initial equilibrium value is
$\overline{\left\langle J_{x,i} \right\rangle} = 2 J_x^{\text{ETH}}
\sum_j \cos \beta_j \left|\Gamma_{j,\pm}\right| $, we conclude that
\begin{equation}
\overline{\left\langle J_{x,f} \right\rangle} =
\frac{\overline{\left\langle J_{x,i} \right\rangle}}{M} \sum_j^M \cos
\omega_j.
\end{equation}

Two important physical conclusions arise from this result. First,
$\left| \overline{\left\langle J_{x,f} \right\rangle} \right| \leq
\left| \overline{\left\langle J_{x,i} \right\rangle} \right|$ for any
initial state satisfying the previous conditions and for any
slow-enough process.  This entails that, if $\sum_j^M \cos(\omega_j) <
M$ for any value of $\tau_q$ and $t_r$, the protocol turns out to be
irreversible, because $\overline{\left\langle J_{x,i} \right\rangle}$
is not recovered at the end. Moreover, if the phases $\omega_j$ are
different for every doublet, we can expect that $\sum_j^M \frac{\cos
  \omega_j}{M} \sim 0$, provided the initial state is spread over a
large enough number of eigenstates of the initial Hamiltonian, as it
happens in the thermodynamic limit. So, in any case a relevant part of
the information about the initial symmetry-breaking, encoded in the
initial value $\overline{\left\langle J_{x,i} \right\rangle}$, is
irreversibly lost after the protocol. For a clear interpretation of
this result, it is worth to have in mind that the choice of the
time-dependent protocol $\Lambda(t)$ constitutes an one degree of
freedom {\em macroscopic} trajectory, which produces a very large
number of different {\em microscopic} phases. Therefore, to exactly
revert the process $H(t) \rightarrow H(-t)$, a Maxwell's demon-like
machine~\cite{Muruyama:09} with a fully microscopic knowledge of the
system is required, exactly as for any other thermodynamic
irreversible phenomenon.

On the other hand, if the initial state only populates one doublet,
the final result is $\overline{\left\langle J_{x,f} \right\rangle} =
\overline{\left\langle J_{x,i} \right\rangle} \cos \omega_m (\tau_q,
t_r)$. This implies that the final equilibrium value
$\overline{\left\langle J_{x,f} \right\rangle}$ oscillates between
$-\overline{\left\langle J_{x,i} \right\rangle}$ and
$\overline{\left\langle J_{x,i} \right\rangle}$, depending on the
precise values of $\tau_q$ and $t_r$. Hence, the information about the
initial symmetry-breaking is not lost at all after the forward
process, as it happens if the initial state populates a large number
of doublets ---the initial value $\overline{\left\langle J_{x,i}
  \right\rangle}$ can be easily recovered selecting $t_r$ in the
normal phase to make $\omega_m (\tau_q,t_r) = 2 \pi n$, being $n=0, 1,
2,\ldots$. The main difference between this and the previous case is
that the protocol gives rise to a single phase $\omega_m (\tau_q,t_r)$
in this one, which can be removed by means of a precise choice of
$t_r$ and $\tau_q$.

For small systems, in which only a small number of doublets are
populated in the initial state, we can have a intermediate result with
$\overline{\left\langle J_{x,f} \right\rangle}$ oscillating between
$-\hat{J}$ and $\hat{J}$. This is what we plot as a band in
Fig.~\ref{fig:cycle}, being $\hat{J}$ smaller as the number of
initially populated doublets is increased, as we show in
Fig.~\ref{fig:band}. The exact value of $\overline{\left\langle
  J_{x,f} \right\rangle}$ depends on the precise values of the phases
$\omega_j$, but we can expect that almost any protocol gives rise to a
quite large number of different phases $\omega_j$, at least if the
system is large enough to have a large number of doublets populated in
the initial state, and hence $\overline{\left\langle J_{x,f}
  \right\rangle} \sim 0$.

\section{Conclusions} 

In this article we report the existence of irreversible processes {\em
  without} energy dissipation in a certain kind of isolated quantum
systems. Starting with a symmetry-breaking state in which only one of
the two symmetry-breaking branches is populated, and changing a system
parameter slowly enough, we perform a closed cycle which is
quasistatic ---the initial energy distribution is perfectly recovered
at the end, and hence the net work necessary to complete the cycle is
zero. However, the process results to be irreversible, because the
final equilibrium state consists of a superposition of the two
symmetry-breaking branches, no matters the fine details of the initial
condition. So, this kind of irreversibility is not related to the
dissipation of energy, but to an unavoidable loss of information,
which is caused by the collective phase mixing between different
symmetry sectors when the system enters in a phase with no
degeneracies. This phase mixing entails the unavoidable erasure of the
information about the initial symmetry-breaking, making the process
irreversible, no matters how slow it is performed. We exemplify this
fact by means of numerical calculations in an experimental feasible
many-body quantum system, which is a special case of the
Lipkin-Meshkov-Glick model. We perform an irreversible cycle
consisting of a unitary time evolution that entails a loss of
information $\Delta I \sim \log 2$ in the measurement of a
symmetry-breaking observable, that we identify with an increase of the
von Neumann entropy of the long-time average equilibrium states, that
tends to $\Delta S = \log 2$ in the quasistatic limit.

\begin{acknowledgments}
 We would like to thank O. Marty for useful discussions. The work is
 supported by Spanish Government grant for the research project
 FIS2012-35316, an Alexander von Humboldt Professorship, the EU
 Integrating Project SIQS and the EU STREP project EQUAM.  Part of the
 calculations of this work were performed in the high capacity cluster
 for physics, funded in part by Universidad Complutense de Madrid and
 in part with Feder funding. This is a contribution to the Campus of
 International Excellence of Moncloa, CEI Moncloa.
\end{acknowledgments}

\end{document}